\documentclass{article}

\usepackage{PRIMEarxiv}

\usepackage[utf8]{inputenc} 
\usepackage[T1]{fontenc}    
\usepackage{hyperref}       
\usepackage{url}            
\usepackage{booktabs}       
\usepackage{amsfonts}       
\usepackage{nicefrac}       
\usepackage{microtype}      
\usepackage{lipsum}
\usepackage{amsmath}
\usepackage{amssymb}
\usepackage{fancyhdr}       
\usepackage{graphicx}       
\usepackage[super]{cite}
\usepackage{xcolor}
\usepackage{float}
\hypersetup{colorlinks=false,allbordercolors=blue,pdfborderstyle={/S/U/W 1}}
\graphicspath{{media/}}     

\title{Constraining Cosmological Parameters From Statistical Superluminal Effects Without a Distance Ladder
\thanks{\textit{} 
\textbf{Accepted for publication in Modern Physics Letters A (2026)
DOI: 10.1142/S0217732326501750}} 
}

\author{
  Hichem Guergouri \\
 Research Unit in Scientific Culture and Mediation - CERIST\\
Constantine, 25016,
Algeria
  \texttt{h.guergouri@cerist.dz} \\
   \And
  Omar Nemoul \\
  Research Unit in Scientific Culture and Mediation - CERIST\\
Constantine, 25016,
Algeria
  \texttt{onemoul@cerist.dz} \\
    \And
  Jamal Mimouni \\
  Research Unit in Scientific Culture and Mediation - CERIST\\
Constantine, 25016,
Algeria\\
Laboratoire de Physique Mathématique et Subatomique (LPMS)\\
University of Constantine 1\\
Constantine, 25017, Algeria
  \texttt{jmimouni@cerist.dz} \\
}

\begin{document}
\maketitle

\begin{abstract}
We employ a statistical approach to study apparent superluminal motion of luminous sources in an expanding flat Friedmann--Lemaître--Robertson--Walker universe, explicitly incorporating cosmological effects through the comoving distance at the emission time. Probability Density Functions (PDFs) of the apparent angular velocity are derived under minimal assumptions regarding source orientations and intrinsic peculiar velocity distributions. We show that the apparent angular velocity distributions and their associated statistical observables are sensitive to cosmological parameters $\Omega_{\Lambda,0}$ and the Hubble parameter $H_0$. Using suitably defined observables, we construct correlated constraints in the $(\Omega_{\Lambda,0}, H_0)$ parameter space and demonstrate that combining measurements at different redshifts effectively breaks the resulting degeneracy. Apparent superluminal motion thus provides a complementary kinematic consistency test for cosmological models.
\end{abstract}


\section{Introduction}\label{sec1}

Apparent superluminal motion is one of the most striking manifestations of special relativity in astrophysics. It refers to the observational phenomenon in which compact plasma features (``blobs'') in relativistic jets appear to move across the sky with transverse velocities exceeding the speed of light, even though their true bulk speed remains strictly subluminal. This effect was first predicted in the theoretical works of Martin J.~Rees in the 1960s~\cite{Rees1966}, who showed that luminous sources moving at relativistic speeds and observed at small angles to the line of sight may naturally appear to a distant observer to have superluminal transverse velocity. A few years later, convincing observational confirmations were reported through very-long-baseline interferometry (VLBI) observations of compact radio sources, most notably in quasars such as 3C~273 and 3C~279~\cite{Whitney1971}, establishing superluminal motion as a real and measurable astrophysical phenomenon. Since these pioneering discoveries, superluminal motion has been extensively studied in the context of active galactic nuclei (AGN), radio galaxies, and blazars (jets oriented close to the line of sight). A comprehensive theoretical framework was developed by Blandford and Rees~\cite{Blandford1974}, and later refined in population studies by Urry and Padovani~\cite{Urry1995}. Modern large-sample surveys, such as those conducted by the MOJAVE collaboration~\cite{Lister2016,Lister2019}, have provided statistically robust measurements of apparent jet speeds and Doppler boosting factors, enabling detailed comparisons between theory and observation across hundreds of relativistic jets.

While most classical treatments of superluminal motion assume a static Minkowski background or a purely special-relativistic framework, realistic astrophysical sources are embedded in an expanding Universe. At cosmological distances, light propagation, observed timescales, and apparent angular motions are all affected by cosmic expansion through the factor $(1+z)^{-1}$, equivalently expressed via the scale factor $a(t)$ at the emission time. This introduces subtle but important corrections to the apparent transverse velocity when the emitting source is located at non-negligible redshift. The cosmological context of relativistic apparent motions has been explored in several works~\cite{Urry1995,Vermeulen1994,Chodorowski2005}, demonstrating that cosmic expansion modifies observed velocities through cosmological time dilation and distance--redshift relations and must be accounted for in precision modeling of high-redshift jets. Closely related to superluminal motion is the phenomenon of Doppler boosting (relativistic beaming), whereby radiation from a source moving at relativistic speed is strongly amplified in the forward direction. The Doppler factor controls the observed flux density, variability timescales, and brightness temperature of relativistic jets, and plays a central role in the observed dominance of blazars in flux-limited surveys (see Ref.~\cite{Urry1995} for a comprehensive review).

Recent studies~\cite{Son2025} have questioned the reliability of Type~Ia supernovae as standardizable candles, showing that their standardized luminosities depend significantly on the age of their progenitor stellar populations. This age dependence, detected at the $5.5\sigma$ level, introduces a redshift-dependent systematic bias that is not properly corrected by the usual host-galaxy mass-step calibration. After accounting for this bias, supernova data no longer favor the standard $\Lambda$CDM model and instead align more closely with the $\mathrm{w}_{0}\mathrm{w}_{a}$CDM model recently suggested by the DESI BAO project from a combined analysis using only BAO and CMB data~\cite{Karim2025}. The combined analysis reveals a strong tension with $\Lambda$CDM and suggests a time-varying dark energy equation of state in a non-accelerating universe. These findings highlight the importance of developing distance-ladder-independent methods for probing and constraining cosmological parameters.

Returning to transverse motions, it is noteworthy that the current observational record for extragalactic jets extends to very high redshifts, as established by VLBI proper-motion measurements of distant blazars and quasars. Representative examples include the blazar J1430+4204 at redshift $z=4.72$~\cite{Zhang2020} and the quasar PMN~J2134--0419 at $z=4.33$~\cite{Perger2018}. Such redshifts correspond to comoving distances of order $\sim20~\mathrm{Gly}$, providing a unique window into relativistic jet kinematics at early cosmological epochs and offering potential constraints on cosmological model parameters.

Motivated by these considerations, the present work develops a statistical framework for modeling the apparent superluminal effect in an expanding cosmological background. Rather than focusing on individual sources, we treat superluminal motion as a population-level phenomenon governed by well-defined probability distributions of jet orientations and intrinsic (peculiar) velocities, while explicitly incorporating the influence of cosmology through the cosmic expansion history at the emission epoch. Throughout this work, the Universe is modeled as a spatially flat Friedmann--Lemaître--Robertson--Walker (FLRW) spacetime~\cite{Friedmann1967,Lemaitre1979,Robertson1935,Walker1937} containing two components: pressureless matter and a cosmological constant, characterized by a dark energy density parameter $\Omega_{\Lambda,0}$ (with $\Omega_{\mathrm{m},0}+\Omega_{\Lambda,0}=1$) and the present-day Hubble constant $H_0$. The impact of cosmological parameters enters explicitly through the comoving distance $\chi$ evaluated at the emission time, thereby linking apparent transverse motion directly to the expansion history of the Universe.

Our objective is to derive semi-analytical probability density functions (PDFs) for the apparent angular velocity $\dot{\phi}_{\mathrm{app}}$ and to investigate how the cosmological parameters $\Omega_{\Lambda,0}$ and $H_0$ control the resulting observable distributions. Furthermore, we aim to construct statistical observables capable of constraining the $(\Omega_{\Lambda,0}, H_0)$ parameter space without reliance on a distance ladder. This work is not intended as an astrophysical study of jet phenomenology or data analysis, but rather as a cosmological modeling framework designed to isolate and quantify the imprint of cosmic expansion on relativistic kinematic observables.

This paper is organized as follows. In Section~\ref{section2}, we introduce the concept of apparent velocity in both static and expanding universes. Section~\ref{section3} derives the general expression for the apparent angular velocity distribution of a population of luminous sources at fixed redshift. In Section~\ref{section4}, we construct a toy model combining intrinsic velocity distributions, isotropic orientations, and cosmological expansion, and provide numerical illustrations for different dispersions and cosmological models. Section~\ref{section5} explores the ability of apparent angular velocity statistics to constrain cosmological parameters without a distance ladder, showing that single-redshift populations lead to strong parameter degeneracies that can be broken by combining measurements across multiple redshifts. Finally, Section~\ref{section6} summarizes our main conclusions. Throughout this work, we adopt units in which the speed of light $c=1$.

\section{Apparent Transverse Motion}\label{section2}

\subsection{Apparent Transverse Motion in Flat and Static Spacetime}
Consider an emitting source (often referred to as a ``blob'' in the context of active galactic nuclei or relativistic jets) moving with intrinsic velocity $v$ at an angle $\theta$ with respect to the observer's line of sight. In the standard special-relativistic treatment, the apparent transverse velocity arises from light-travel--time effects and is given by

\begin{equation}
v_{\text{app}} = \frac{v \sin\theta}{1 - v \cos\theta}.
\label{eq1}
\end{equation}
The derivation of Eq.~\eqref{eq1} can be found in many textbooks on astrophysics (see, e.g., Refs~\cite{Krolik1999,Shu1982}). This expression shows that a source can appear to the observer to move transversely at subluminal, luminal, or even superluminal speeds without violating causality. The apparent regime depends on the viewing angle $\theta$ and the intrinsic velocity $v$. Consequently, the ($\theta,v$) parameter space can be partitioned into distinct regions corresponding to these three cases, with the luminal boundary separating the subluminal and superluminal domains. It is worth emphasizing that the quantity directly observable is the apparent angular velocity, $\dot{\phi}_{\mathrm{app}}$. The apparent transverse velocity is then inferred by multiplying this angular rate by the distance to the source at the time of emission, yielding $v_{\mathrm{app}} = \chi(t_\text{emit}) \, \dot{\phi}_{\mathrm{app}}$ where $\chi(t_\text{emit})$ denotes the distance to the source at the emission time.

\subsection{Apparent Transverse Motion in FLRW Spacetime}
Now, for sources located at cosmological distances, the above relation~\eqref{eq1} must be modified to account for the expanding spacetime. In a spatially flat Friedmann--Lemaître--Robertson--Walker (FLRW) spacetime, physical transverse distances are related to comoving distances through the scale factor $a(t)$. At the emission time $t_{\text{emit}}$, the physical transverse displacement is reduced by a factor $a(t_{\text{emit}})/a(t_{\text{obs}})$ relative to its present-day comoving value at $t_{\text{obs}}$. As a result, the apparent transverse velocity measured by the observer is reduced by the same factor, equivalently, divided by $1+z$, where
$z$ is the cosmological redshift of the source. This yields the following cosmological expression
\begin{equation}
v_{\text{app}} =\frac{a(t_{\text{emit}})}{a(t_{\text{obs}})} \frac{v \sin\theta}{1 - v \cos\theta}.
\label{eq2}
\end{equation}
Here, the apparent peculiar velocity $v_\text{app}=a(t_\text{emit})\chi(t_\text{emit})\dot{\phi}_{\mathrm{app}}$. The factor $a(t_{\mathrm{emit}})/a(t_{\mathrm{obs}}) = (1+z)^{-1}$ corresponds to the standard cosmological time-dilation effect; for large distances between the observer and the source, the emission time $t_\text{emit}$ is earlier, corresponding to a smaller scale factor $a(t_{\text{emit}})$. As a result, the apparent transverse velocity $v_{\text{app}}$ appears smaller than it would in a static universe.

To rigorously derive Eq.~\eqref{eq2}, for sources located at cosmological distances, we begin by considering a variation of the comoving distance,
\begin{equation}
\chi(t_{\mathrm{emit}}) = \int_{t_{\mathrm{emit}}}^{t_{\mathrm{obs}}} \frac{dt}{a(t)} \, .
\label{eq3}
\end{equation}
Taking a first-order variation of this expression with respect to the emission and observation times, one finds
\begin{equation}
\frac{dt_{\mathrm{emit}}}{dt_{\mathrm{obs}}}
=
\frac{a(t_{\mathrm{emit}})/a(t_{\mathrm{obs}})}
{1 + a(t_{\mathrm{emit}})\,\dot{\chi}(t_{\mathrm{emit}})} \, ,
\label{eq4}
\end{equation}
where $\dot{\chi} \equiv d\chi/dt$ denotes the time derivative of the comoving distance. The apparent angular velocity is defined by relating the angular variation at emission to the observer’s local time,
\begin{equation}
\begin{aligned}
\dot{\phi}_{\mathrm{app}}
&=
\frac{\phi(t_{\mathrm{emit}} + dt_{\mathrm{emit}}) - \phi(t_{\mathrm{emit}})}{dt_{\mathrm{obs}}} \\
&=
\dot{\phi}(t_{\mathrm{emit}})\,
\frac{dt_{\mathrm{emit}}}{dt_{\mathrm{obs}}} .
\end{aligned}
\label{eq5}
\end{equation}
Using the previous result in Eq.~\eqref{eq4}, this can be written as
\begin{equation}
\dot{\phi}_{\mathrm{app}}
=
\frac{1}{a(t_{\mathrm{obs}})}
\frac{a(t_{\mathrm{emit}})\,\dot{\phi}(t_{\mathrm{emit}})}
{1 + a(t_{\mathrm{emit}})\,\dot{\chi}(t_{\mathrm{emit}})} \, .
\label{eq6}
\end{equation}
We now define the peculiar velocity of the source as
\begin{equation}
v_{\mathrm{pec}}(t)
=
\sqrt{v_{\mathrm{pec},\chi}^2(t) + v_{\mathrm{pec},\phi}^2(t)} \, ,
\qquad
0 \le v_{\mathrm{pec}}(t) < 1 \, ,
\label{eq7}
\end{equation}
with the radial and transverse components given by
\begin{subequations}
\begin{align}
v_{\mathrm{pec},\chi}(t) &= a(t)\,\dot{\chi}(t)\label{eq8a}, \\
v_{\mathrm{pec},\phi}(t) &= a(t)\,\chi(t)\,\dot{\phi}(t).
\end{align}
\label{eq8b}
\end{subequations}
Here, $a(t)\chi(t)$ is the angular diameter distance $D_A(z)$, and $\chi(t)$ coincides with the metric (comoving) distance\footnote{For a spatially curved spacetime with constant curvature $k$, the metric distance is given by
$\frac{1}{\sqrt{k}} \sin\!\left(\sqrt{k}\,\chi\right)$ for a spherical geometry ($k>0$), and by
$\frac{1}{\sqrt{|k|}} \sinh\!\left(\sqrt{|k|}\,\chi\right)$ for a hyperbolic geometry ($k<0$),
where $\chi$ denotes the comoving distance, which coincides with the metric distance in the flat case $k=0$.} in a spatially flat universe. For $\Lambda$CDM model, the non-monotonic behavior of the angular diameter distance $D_A(z)$, which causes very distant objects (beyond $z \sim 1.6$) to appear larger in angular size, potentially enhancing apparent angular velocities. Substituting these definitions into the expression for $\dot{\phi}_{\mathrm{app}}$, we obtain
\begin{equation}
\dot{\phi}_{\mathrm{app}}
=
\frac{1}{a(t_{\mathrm{obs}})}\,
\frac{1}{\chi(t_{\mathrm{emit}})}\,
\frac{v_{\text{pec},\phi}(t_{\mathrm{emit}})}
{1 + v_{\text{pec},\chi}(t_{\mathrm{emit}})} \, .
\label{eq9}
\end{equation}
Introducing the parametrization of the peculiar velocity components, $v_{\text{pec},\chi}(t_{\mathrm{emit}}) = - v \cos\theta$ and $v_{\text{pec},\phi}(t_{\mathrm{emit}}) = v \sin\theta$, where $v=v_\text{pec}(t_\text{emit})$ and $\theta$  is the angle between the peculiar velocity vector and the observer’s line of sight in the comoving frame, leads to
\begin{equation}
\dot{\phi}_{\mathrm{app}}
=
\frac{1}{a(t_{\mathrm{obs}})}\,
\frac{1}{\chi(t_{\mathrm{emit}})}\,
\frac{v \sin\theta}{1 - v \cos\theta} \, .
\label{eq10}
\end{equation}
Finally, using the relation $v_{\mathrm{app}} = a(t_{\mathrm{emit}})\,\chi(t_{\mathrm{emit}})\,\dot{\phi}_{\mathrm{app}}$, one recovers the familiar expression given in Eq.~\eqref{eq2}. In the following, we fix the scale factor at the present (observation) time, setting $a(t_{\text{obs}})=1$.

\section{Statistical Modeling of Apparent Angular Velocity}\label{section3}

In this framework, we consider a population of emitting sources located at the same redshift $z$, so that they share a common comoving distance $\chi(t_{\mathrm{emit}})$. Each source is characterized by: \textit{(i)} an orientation angle $\theta \in [0,\pi]$ specifying its direction of motion relative to the observer, with the azimuthal angle $\phi$ uniformly distributed for each $\theta$ due to axial symmetry, and \textit{(ii)} an intrinsic peculiar velocity $v\in\left[\left.0;1\right)\right.$ relative to the comoving frame centered on the observer. We assume that the statistical distributions of their directional orientations $P_{\theta}(\theta)$ and peculiar velocities $P_v(v)$ are well characterized and can be described by well-defined probability laws, depending on the physical nature of the phenomenon under study.

To derive the apparent angular velocity PDF, it is convenient to
express $\dot{\phi}_{\mathrm{app}}$ from Eq.~\eqref{eq10} as a multi-variable function  $f(\theta, v)$, which is given by
\begin{equation}
f(\theta, v)
= \frac{1}{\chi(t_{\mathrm{emit}})}\,\frac{ v \sin\theta}{1 - v \cos\theta}
= \dot{\phi}_{\mathrm{app}},
\label{eq11}
\end{equation}
and its inverse function $g(\theta, \dot{\phi}_{\mathrm{app}})$ with respect to the peculiar velocity $v$
\begin{equation}
g(\theta, \dot{\phi}_{\mathrm{app}})
= \frac{\chi(t_{\mathrm{emit}})\,\dot{\phi}_{\mathrm{app}}}
       {\sin\theta + \chi(t_{\mathrm{emit}})\dot{\phi}_{\mathrm{app}}\cos\theta}\,
= v .
\label{eq12}
\end{equation}
These relations fully encode the transformation between intrinsic and apparent
kinematic quantities at a given emission time $t_{\text{emit}}$.
The probability density function of the apparent angular velocity is obtained from the
joint distribution of $v$ and $\theta$ as
\begin{equation}
P_{\dot{\phi}_{\mathrm{app}}}(\dot{\phi}_{\mathrm{app}})
= \int_{0}^{\pi} d\theta \int_{0}^{1} dv\;
P_{v}(v)\,P_{\theta}(\theta)\,
\delta\!\left( \dot{\phi}_{\mathrm{app}} - f(\theta,v) \right).
\label{eq13}
\end{equation}
Throughout this analysis, the variables $\theta$ and $v$ are treated as statistically independent. 
Equation~\eqref{eq13} shows that, for a fixed value of the apparent angular velocity $\dot{\phi}_{\mathrm{app}}$, 
the corresponding probability density is obtained by integrating the joint weight
$P(\theta,v)=P_{\theta}(\theta)P_v(v)$ over all points $(\theta,v)$ lying on the curve $\gamma_{\dot{\phi}_{\mathrm{app}}}:\quad f(\theta,v)=\dot{\phi}_{\mathrm{app}}$ in the $(\theta,v)$ plane. Using the standard change of variables formula for the Dirac delta distribution, we may rewrite
\begin{equation}
\delta\!\left(\dot{\phi}_{\mathrm{app}} - f(\theta, v)\right)
=
\frac{\chi(t_{\mathrm{emit}})\,\sin\theta}
{\left[\sin\theta + \chi(t_{\mathrm{emit}})\dot{\phi}_{\mathrm{app}}\cos\theta\right]^{2}}
\,
\delta\!\left(v - g(\theta, \dot{\phi}_{\mathrm{app}})\right),
\label{eq14}
\end{equation}
Substituting this expression into Eq.~\eqref{eq13} yields the semi-analytical form
\begin{equation}
P_{\dot{\phi}_{\mathrm{app}}}(\dot{\phi}_{\mathrm{app}})
= \int_{\gamma_{\dot{\phi}_{\mathrm{app}}}} d\theta\;
\frac{\chi(t_{\mathrm{emit}})\,\sin\theta\,\,P_{v}\!\left( g(\theta,\dot{\phi}_{\mathrm{app}}) \right)\,
P_{\theta}(\theta)}
{\left[\sin\theta + \chi(t_{\mathrm{emit}})\dot{\phi}_{\mathrm{app}}\cos\theta\right]^{2}}.
\label{eq15}
\end{equation}
where the restricted $\theta$-integration occurs along the curve $\gamma_{\dot{\phi}_{\mathrm{app}}}$ for a fixed $\dot{\phi}_{\mathrm{app}}$ in the ($\theta,v$) plane. Since all curves $\gamma_{\dot{\phi}_{\mathrm{app}}}$ of constant
$\dot{\phi}_{\mathrm{app}}$ in the $(\theta,v)$ plane originate at
$(\theta=0,\,v=1)$, initially decrease to reach a local minimum at
$(\theta_0,\,v_0)$, and then increase again, finally terminating at
$(\theta_\ast,\,v=1)$ (see the color plots in Figs.~\ref{fig3}), the $\theta$-integration domain is bounded by
$0 < \theta < \theta_\ast$, where the upper limit $\theta_{*}$ is defined
implicitly as the nonzero solution of the equation $g(\theta_{*}, \dot{\phi}_{\mathrm{app}}) = 1$. This leads to the relations
\begin{subequations}
\begin{align}
\sin\theta_{*}
&=
\frac{2 \chi(t_{\mathrm{emit}})\,\dot{\phi}_{\mathrm{app}}}
     {\chi^2(t_{\mathrm{emit}})\,\dot{\phi}_{\mathrm{app}}^{2} + 1}
\equiv s_{1}(\dot{\phi}_{\mathrm{app}}),
\label{eq16a} \\[0.5ex]
\cos\theta_{*}
&=
\frac{\chi^2(t_{\mathrm{emit}})\,\dot{\phi}_{\mathrm{app}}^{2} - 1}
     {\chi^2(t_{\mathrm{emit}})\,\dot{\phi}_{\mathrm{app}}^{2} + 1}
\equiv c_{1}(\dot{\phi}_{\mathrm{app}}),
\label{eq16b}
\end{align}
\end{subequations}
and therefore $\theta_{*}$ as a function of $\dot{\phi}_{\mathrm{app}}$ is given by
\begin{equation}
\theta_{*}(\dot{\phi}_{\mathrm{app}})
= \operatorname{atan2}\!\left( s_{1}(\dot{\phi}_{\mathrm{app}}),\; c_{1}(\dot{\phi}_{\mathrm{app}}) \right).
\label{eq17}
\end{equation}
Throughout this work, velocities are expressed in units of the speed of light $c$, while the Hubble constant $H_0$ is adopted in units of $\mathrm{Gyr}^{-1}$. Accordingly, the comoving distance $\chi(t_{\mathrm{emit}})$ is expressed in gigalight-years ($\mathrm{Gly}$). The resulting unit of angular velocity is therefore $c/\mathrm{Gly}$, which can be converted into the observational unit of milliarcseconds per year ($\mathrm{mas}\,\mathrm{yr}^{-1}$) as $1\,c/\mathrm{Gly} = \frac{81}{125\pi}\,\mathrm{mas}\,\mathrm{yr}^{-1}
\simeq 0.2\,\mathrm{mas}\,\mathrm{yr}^{-1} $. Using the corresponding Jacobian for the change of units, the full expression of the apparent angular velocity (in $\mathrm{mas}\,\mathrm{yr}^{-1}$) PDF becomes
\begin{equation}
P_{\dot{\phi}_{\mathrm{app}}}(\dot{\phi}_{\mathrm{app}})
=
\int_{0}^{\theta_{*}(\frac{125\pi}{81}\dot{\phi}_{\mathrm{app}})} d\theta\;
\frac{\frac{125\pi}{81}\chi(t_\text{emit})\,\sin\theta\,\,\,P_{v}\!\left( g(\theta,\frac{125\pi}{81}\dot{\phi}_{\mathrm{app}}) \right)\,
P_{\theta}(\theta)}
            {\left[\sin\theta + \chi(t_{\mathrm{emit}})\frac{125\pi}{81}\dot{\phi}_{\mathrm{app}}\cos\theta\right]^{2}}.
\label{eq18}
\end{equation}
This normalized expression provides a fully general formulation valid for any choice of intrinsic velocity distribution $P_v(v)$ and orientation distribution $P_\theta(\theta)$. It therefore provides a flexible framework within which alternative statistical models can be readily implemented. This completes the formal construction of the semi-analytical PDF of the apparent angular velocity. In the next section, we specialize this general result to isotropically distributed orientations and a logit-normal velocity distribution, and analyze the resulting apparent angular velocity distributions.

\section{Illustrative Toy Model}\label{section4}
\subsection{Orientation Distribution}
In the absence of preferred cosmic directions, the intrinsic orientations of the luminous sources under study are expected to be random. For a sufficiently large sample, the orientation distribution can therefore be considered isotropic. For the numerical illustrations of Eq.~\eqref{eq18}, we adopt an isotropic orientation distribution as a toy model:
\begin{equation}
P_{\theta}(\theta) = \frac{1}{2}\,\sin\theta.
\label{eq19}
\end{equation}
In practice, observations of relativistic jets and blobs do not show a perfectly uniform orientation distribution. Empirical data indicate a strong bias toward jets oriented close to the line of sight, primarily due to Doppler boosting: jets pointing toward the observer appear significantly brighter because of relativistic beaming (see, e.g.,~\cite{Urry1995,Lister2016,Lister2019}). Instrumental selection effects associated with finite telescope sensitivity further enhance this preference for bright, forward-facing sources.

The formalism developed in this section is sufficiently general to allow modified Doppler boosting prescriptions and alternative empirical or physically motivated orientation models, enabling future extensions of the statistical description of apparent angular velocities.

\subsection{Intrinsic (Peculiar) Velocity Distribution}
A key ingredient of the model is the statistical description of intrinsic (peculiar) velocities. 
A simple Gaussian distribution in the velocity variable $v$ is not appropriate, since peculiar velocities are physically bounded by
$0 \le v < 1$, whereas a Gaussian distribution extends to $\pm\infty$ and assigns nonzero probability to unphysical velocities approaching or exceeding the speed of light. 
It is therefore more natural to adopt a probability distribution explicitly defined on a bounded domain.
A convenient choice is the \emph{logit-normal} distribution. 
We introduce an auxiliary variable $u$ through the bijective mapping $u:(0,1)\rightarrow(-\infty,+\infty)$,
\begin{equation}
u(v) = \ln\!\left(\frac{v}{1-v}\right).
\label{eq20}
\end{equation}
The transformed variable $u$ is assumed to follow a normal distribution,
$u \sim \mathcal{N}(\mu_u,\sigma_u^2)$, with Gaussian probability density $P_u(u)$.
The induced distribution for $v$ then satisfies
$P_v(v)\,dv = P_u(u)\,du$.
Using
$\frac{du}{dv} = [v(1-v)]^{-1}$,
one obtains the logit-normal distribution
\begin{equation}
P_v(v)
=
\frac{1}{\sigma_u\sqrt{2\pi}}
\frac{1}{v(1-v)}
\exp\!\left[
-\frac{1}{2\sigma_u^2}
\left(
\ln\!\left(\frac{v}{1-v}\right)-\mu_u
\right)^2
\right].
\label{eq21}
\end{equation}
The logit-normal distribution provides a Gaussian-like description for a bounded variable and reduces to an approximately Gaussian form for sufficiently small $\sigma_u$. 
For physical transparency, it is desirable to reparametrize the distribution in terms of the most probable velocity $v_{\mathrm{peak}}$ and the velocity dispersion $\sigma_v$.
The location parameter $\mu_u$ is determined by requiring that the distribution peaks at $v=v_{\mathrm{peak}}$, namely
$\left.\frac{dP_v}{dv}\right|_{v=v_{\mathrm{peak}}}=0$.
This condition yields
\begin{equation}
\mu_u
=
\ln\!\left(\frac{v_{\mathrm{peak}}}{1-v_{\mathrm{peak}}}\right)
-
\sigma_u^2\left(2v_{\mathrm{peak}}-1\right).
\label{eq22}
\end{equation}
In the limit $\sigma_u \rightarrow 0$, this expression reduces to
\begin{equation}
\mu_u
\approx
\ln\!\left(\frac{v_{\mathrm{peak}}}{1-v_{\mathrm{peak}}}\right),
\label{eq23}
\end{equation}
and the distribution $P_v(v)$ becomes sharply peaked around $v \approx v_{\mathrm{peak}}$.
In this regime, a linear expansion of the logarithmic term appearing in the exponent of Eq.~\eqref{eq21} about $v=v_{\mathrm{peak}}$ gives
\begin{equation}
\ln\!\left(\frac{v}{1-v}\right)-\mu_u
\approx
\frac{v-v_{\mathrm{peak}}}{v_{\mathrm{peak}}(1-v_{\mathrm{peak}})}.
\label{eq24}
\end{equation}
Substituting this approximation into the distribution~\eqref{eq21}, one finds that the
logit-normal distribution reduces locally to a Gaussian distribution in the
peculiar velocity,
$v \sim \mathcal{N}(v_{\mathrm{peak}}, \sigma_v^{2})$.
In this limit, the dispersion parameters are related through
\begin{equation}
\sigma_u
\approx
\frac{\sigma_v}{v_{\mathrm{peak}}(1-v_{\mathrm{peak}})}.
\label{eq25}
\end{equation}
Throughout our analysis we consider small velocity dispersions, $\sigma_v \ll 1$. 
For typical values of the most probable velocity $v_{\mathrm{peak}}$ away from the boundaries,
this immediately implies $\sigma_u \ll 1$, ensuring that Eq.~\eqref{eq25} provides an accurate approximation
for the dispersion $\sigma_u$ in terms of $\sigma_v$. For values of $v_{\mathrm{peak}}$ approaching the boundaries $v_{\mathrm{peak}}\approx 0$ or $v_{\mathrm{peak}}\approx 1$,
undesirable boundary effects (non-Gaussian distortions) can be avoided by imposing the condition $\sigma_v \le \min\!\left(\frac{1-v_{\mathrm{peak}}}{t},\,\frac{v_{\mathrm{peak}}}{t}\right)$, where $t>0$ controls the distance $t\sigma_v$ relative to the separation between $v_{\mathrm{peak}}$ and the nearest boundary.
Choosing $t\ge3$ ensures that boundary effects are negligible and that the approximate relation~\eqref{eq25} remains an accurate representation of the true standard deviation. Collecting Eqs.~\eqref{eq21},~\eqref{eq22}, and~\eqref{eq25}, the peculiar velocity distribution can be expressed directly in terms of
$v_{\mathrm{peak}}$ and $\sigma_v$ as
\begin{equation}
\begin{aligned}
&P_v(v)
=
\frac{1}{\sigma_v\sqrt{2\pi}}
\frac{v_{\mathrm{peak}}(1-v_{\mathrm{peak}})}{v(1-v)}
\\ &\exp\!\left[
-\frac{v_{\mathrm{peak}}^2(1-v_{\mathrm{peak}})^2}{2\sigma_v^2}
\left(
\ln\!\left(\frac{v}{1-v}\right)
-
\ln\!\left(\frac{v_{\mathrm{peak}}}{1-v_{\mathrm{peak}}}\right)
+
\frac{\sigma_v^2(2v_{\mathrm{peak}}-1)}
{v_{\mathrm{peak}}^2(1-v_{\mathrm{peak}})^2}
\right)^2
\right].
\end{aligned}
\label{eq26}
\end{equation}
In the limit $\sigma_v \rightarrow 0$, this distribution reduces to a perfectly Gaussian
$v \sim \mathcal{N}(\mu_v,\sigma_v^2)$, as expected.
Alternative choices for the peculiar velocity distribution may be adopted in specific physical contexts, such as a log-normal distribution for the logarithmic Lorentz factor $\ln\gamma=-\ln{(1-v^2)}/2$ or a normal distribution for the rapidity $\omega=\tanh^{-1}(v)$. Such parametrizations, however, become ill behaved in the
non-relativistic regime $v \ll 1$. By contrast, the logit-normal distribution offers a flexible and general framework that remains well defined and approximately Gaussian-shaped over the full physical
domain $0\le v<1$.

\subsection{Cosmological Model Framework}
The large-scale dynamics of the Universe are described by the
Friedmann--Lemaître--Robertson--Walker (FLRW) metric, in which the expansion
is governed by the Friedmann equations~\cite{Friedmann1967,Lemaitre1979,Robertson1935,Walker1937}. The energy content of the Universe at the current epoch is
conventionally parametrized by the dimensionless density parameters
\begin{equation}
\Omega_{\text{r},0} = \frac{\rho_{\text{r},0}}{\rho_{\mathrm{crit,0}}}, \qquad
\Omega_{\text{m},0} = \frac{\rho_{\text{m},0}}{\rho_{\mathrm{crit,0}}}, \qquad
\Omega_{\Lambda,0} = \frac{\rho_{\Lambda}}{\rho_{\mathrm{crit,0}}}, \qquad
\Omega_{\text{k},0} = -\,\frac{k}{a_{0}^{2} H_0^{2}},
\label{eq27}
\end{equation}
where $\rho_{\mathrm{r},0}$, $\rho_{\mathrm{m},0}$, and $\rho_{\lambda}$ denote the present-day energy densities of radiation, matter, and dark
energy, respectively. The critical density today is defined as
$\rho_{\mathrm{crit},0} = 3H_0^{2}/(8\pi G)$, where $H_0$ is the present-day
Hubble constant, and $k$ denotes the spatial curvature index. Observationally, these parameters are constrained by a combination of Type~Ia
supernovae~\cite{Riess1998,Perlmutter1999}, baryon acoustic oscillation~\cite{Eisenstein2005}, and cosmic microwave background
measurements~\cite{Aghanim2020}. A persistent tension remains between early-Universe and
late-Universe determinations of the Hubble constant~\cite{Aghanim2020}, commonly referred to as
the Hubble tension~\cite{Wong2020,Verde2019}, with local distance-ladder measurements~\cite{Riess2016,Riess2022} favoring
larger values of $H_0$ than those inferred from the cosmic microwave background. In this toy model, we consider a spatially flat
matter--cosmological constant cosmology. In this case, the Friedmann equation admits an
analytic solution for the cosmic scale factor, which can be written as
\begin{equation}
a(t) = 
\left( \frac{1 - \Omega_{\Lambda,0}}{\Omega_{\Lambda,0}} \right)^{1/3}
\sinh^{2/3}\!\left( \frac{3\sqrt{\Omega_{\Lambda,0}}}{2}\,H_0t \right),
\label{eq28}
\end{equation}
where the Friedmann equation imposes the constraint $\Omega_{\text{m},0}+\Omega_{\Lambda,0}=1$, with $\Omega_{\Lambda,0}$ taking values in the open interval $(0,1)$. The limiting cases $\Omega_{\Lambda,0} \to 0$ and $\Omega_{\Lambda,0} \to 1$ correspond to a matter-dominated (MD) and a dark energy dominated ($\Lambda$D) universe, respectively, while $\Omega_{\Lambda,0} = 0.685$ represents the standard $\Lambda$CDM model. After specifying the cosmological background, the
comoving distance $\chi$ between the luminous source and the observer at the emission time is given by Eq.~\eqref{eq3}. Using the relation $a = 1/(1+z)$ together with the Friedmann equation, the comoving distance can be expressed as
\begin{equation}
\chi(t_{\mathrm{emit}}) = \frac{1}{H_0} \int_0^{z} \frac{dz'}{\sqrt{(1-\Omega_{\Lambda,0}) (1+z')^3 + \Omega_{\Lambda,0}}} \, .
\label{eq29}
\end{equation}
Thus, for fixed $(z,\Omega_{\Lambda,0}, H_0)$, the cosmoving distance $\chi(t_{\mathrm{emit}})$ is
uniquely determined.
Having specified the statistical distributions of source orientations and intrinsic velocities, and established the underlying cosmological framework, we now proceed to a numerical investigation of Eq.~\eqref{eq18} to illustrate the influence of cosmological parameters on the apparent angular velocity distribution.

\subsection{Apparent Angular Velocity Distribution in the Toy Model}
As a concrete model, we adopt the isotropic angular distribution of~\eqref{eq19} together with the logit-normal peculiar velocity distribution of~\eqref{eq26}. Substituting these choices into the general expression for the apparent angular velocity PDF~\eqref{eq18} gives
\begin{equation}
\begin{split}
P_{\dot{\phi}_{\mathrm{app}}}(\dot{\phi}_{\mathrm{app}})
={}& \int_{0}^{\theta_{*}\!\left(\frac{125\pi}{81}\dot{\phi}_{\mathrm{app}}\right)} d\theta\;
\frac{\dfrac{v_{\mathrm{peak}}(1-v_{\mathrm{peak}})}{2\sqrt{2\pi}\sigma_v}\,\,
      \dfrac{\sin^{2}\theta}{\dot{\phi}_{\mathrm{app}}}}
     {\sin\theta - \frac{125\pi}{81}(1-\cos\theta)\chi(t_{\mathrm{emit}})\dot{\phi}_{\mathrm{app}}}
\\[6pt]
&\exp\!\left\{
-\frac{v_{\mathrm{peak}}^{2}(1-v_{\mathrm{peak}})^{2}}{2\sigma_v^{2}}
\left[
\ln\!\left(
\frac{\frac{125\pi}{81}\chi(t_{\mathrm{emit}})\dot{\phi}_{\mathrm{app}}}
     {\sin\theta - \frac{125\pi}{81}(1-\cos\theta)\chi(t_{\mathrm{emit}})\dot{\phi}_{\mathrm{app}}}
\right)
\right.\right.
\\[6pt]
&\left.\left.\,\,\,\,\,\,\,\,\,\,\,\,\,
-\ln\!\left(\frac{v_{\mathrm{peak}}}{1-v_{\mathrm{peak}}}\right)
+\frac{\sigma_v^{2}(2v_{\mathrm{peak}}-1)}
       {v_{\mathrm{peak}}^{2}(1-v_{\mathrm{peak}})^{2}}
\right]^{2}
\right\}.
\end{split}
\label{eq30}
\end{equation}
This expression shows how the logit-normal peculiar velocity distribution characterized by $(v_{\text{peak}},\sigma_v)$, the assumed isotropic orientation distribution, and the cosmoving distance $\chi(t_{\mathrm{emit}})$ through $(z,\Omega_{\Lambda,0},H_0)$ combine to shape the resulting apparent angular velocity distribution. It is semi-analytical in the sense that the $\theta$--integration must generally be performed numerically, while all other dependencies appear explicitly.
To visualize the behavior of the apparent angular velocity PDFs, we numerically evaluate the $\theta$-integral in Eq.~\eqref{eq30} for a population of luminous sources located at a fixed redshift $z = 1$. We adopt an intrinsic velocity distribution characterized by $v_{\text{peak}} = 0.7$\footnote{The adopted value $v_{\text{peak}} = 0.7$ is chosen for illustrative clarity, and that more realistic values can be straightforwardly implemented within the same formalism.}, and three values of the logit-normal dispersion $\sigma_v$, specifically $\sigma_v = 10^{-3}$, $10^{-2}$ and $10^{-1}$. The Hubble parameter is set to $H_0 = 1/14.5\;\mathrm{Gyr}^{-1}$. With these parameters fixed, we numerically evaluate and plot the resulting apparent angular velocity PDFs in observational units of $\mathrm{mas}\,\mathrm{yr}^{-1}$ for three background cosmologies (MD, $\Lambda$CDM, and $\Lambda$D). The resulting PDFs are shown in Figs.~\ref{fig1}.
\subsection*{Discussion}
We now proceed to discuss the main conclusions that can be drawn from these numerical results:
\begin{figure}[]
\centering
\includegraphics[width=0.7\textwidth]{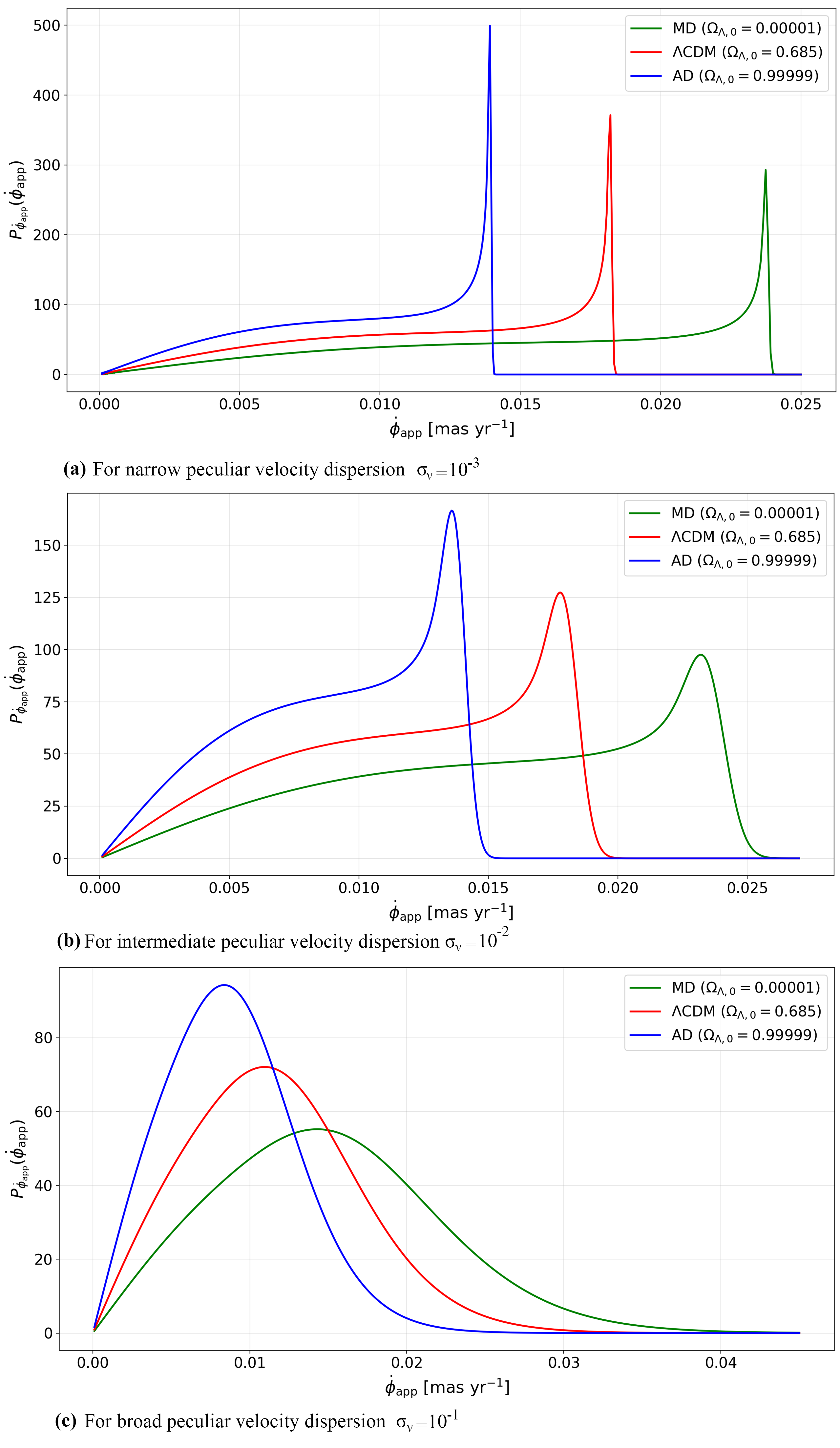}
\caption{Probability density functions of the apparent angular velocity, $P_{\dot{\phi}_{\mathrm{app}}}$, as a function of $\dot{\phi}_{\mathrm{app}}$ for a fixed population of luminous sources at redshift $z=1$. The most probable peculiar velocity is set to $v_{\mathrm{peak}}=0.7$. The present-day Hubble parameter is $H_0 = 1/14.5\,\mathrm{Gyr}^{-1}$. Results are shown for three cosmological models:
matter-dominated (MD; green), $\Lambda$CDM (red), and dark-energy--dominated ($\Lambda$D; blue).}
\label{fig1}
\end{figure}
\begin{itemize}
    \item From Figs.~\ref{fig1}, one can clearly observe that as $\Omega_{\Lambda,0}$ increases from $0$ (matter-dominated case, green curves) to intermediate values ($\Lambda$CDM, red curves), and finally to $\Omega_{\Lambda,0} = 1$ (dark-energy-dominated case, blue curves), the distributions become less dispersed, more localized, and progressively shift toward lower values of $\dot{\phi}_{\mathrm{app}}$. This behavior indicates that both the mean value $\langle \dot{\phi}_{\mathrm{app}} \rangle$ and the standard deviation $\Delta \dot{\phi}_{\mathrm{app}}$ decrease as $\Omega_{\Lambda,0}$ increases. This behavior can be understood physically as follows. If the Universe is entirely dominated by dark energy, i.e., $\Omega_{\Lambda,0} = 1$ (de Sitter spacetime), the expansion rate is maximal and accelerates exponentially. This strong expansion effectively suppresses apparent superluminal effects, leading to a distribution shifted toward lower values of $\dot{\phi}_{\mathrm{app}}$. Conversely, if $\Omega_{\Lambda,0} = 0$, the repulsive effect of dark energy is absent and the expansion is slower. In this case, apparent superluminal motions are less suppressed, and the distribution is shifted toward higher values of $\dot{\phi}_{\mathrm{app}}$ compared to models with larger $\Omega_{\Lambda,0}$.
   \item In contrast, Figs.~\ref{fig2} show that $\langle \dot{\phi}_{\mathrm{app}} \rangle$ and $\Delta \dot{\phi}_{\mathrm{app}}$ increase with the present-day Hubble parameter $H_0$.
   \begin{figure}[!htbp]
   \centerline{\includegraphics[width=0.7\textwidth]{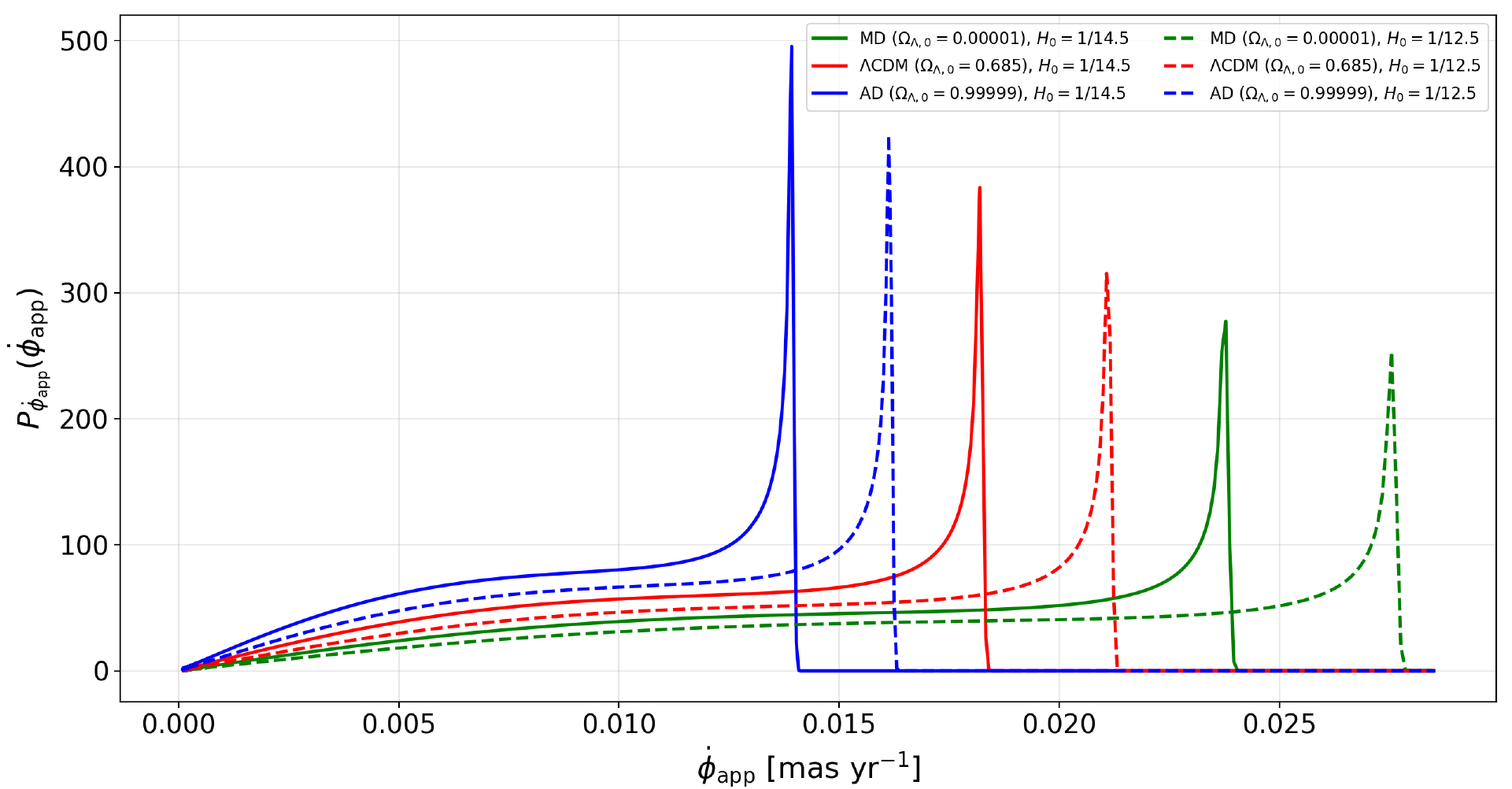}}
    \caption{Probability density functions of the apparent angular velocity,
$P_{\dot{\phi}_{\mathrm{app}}}$, as a function of the apparent angular velocity
$\dot{\phi}_{\mathrm{app}}$ for a luminous source population located at a redshift
$z = 1$. The peculiar velocity distribution is characterized by
$(v_{\mathrm{peak}} = 0.7,\ \sigma_v = 10^{-3})$.
Results are shown for three cosmological models:
matter-dominated (MD; green), $\Lambda$CDM (red),
and dark-energy--dominated ($\Lambda$D; blue).
Solid curves correspond to $H_0 = 1/14.5\,\mathrm{Gyr}^{-1}$,
while dashed curves correspond to $H_0 = 1/12.5\,\mathrm{Gyr}^{-1}$.}
    \label{fig2}
    \end{figure}
    \item For large values of $\sigma_v$ (See Fig.~\ref{fig3}(c) for $\sigma_v = 10^{-1}$), the distribution of apparent angular velocities is unimodal with a right-skewed shape, peaking at $\dot{\phi}_{\mathrm{peak,1}}$. As the velocity dispersion $\sigma_v$ decreases (from $\sigma_v = 10^{-1}$ to $\sigma_v = 10^{-3}$, see Figs.~\ref{fig3}(a,b), the original peak shifts toward higher values, while a second peak $\dot{\phi}_{\mathrm{peak,2}}$ gradually emerges at larger $\dot{\phi}_{\mathrm{app}}$, with $\dot{\phi}_{\mathrm{peak,1}} < \dot{\phi}_{\mathrm{peak,2}}$, and the first peak progressively diminishes in prominence. \item This behavior can be understood as follows: for large $\sigma_v$ (e.g. $\sigma_v = 10^{-1}$, the color plot of Fig.~\ref{fig3}(c)), all points in the $(\theta, v)$ plane contribute significantly to the apparent angular velocity PDF. In this regime, the peak at $\dot{\phi}_{\mathrm{peak,1}}$ arises from a competition of two contributions (i) The effective length of the curve $\dot{\phi}_{\mathrm{app}} = f(\theta,v)$ in the $(\theta,v)$ plane. (ii) The density of intersections between the curve $\dot{\phi}_{\mathrm{app}} = f(\theta,v)$ and regions of high probability in the $(\theta,v)$ color distribution (from purple to yellow). This balance produces the dominant first peak $\dot{\phi}_{\mathrm{peak,1}}$ observed for relatively large dispersion. When $\sigma_v$ decreases, the probability distribution becomes increasingly localized around $v = v_{\mathrm{peak}}$ (yellow horizontal line in the color plot of Fig.~\ref{fig3}(a)). In this regime, contributions from the full $(\theta,v)$ domain become less relevant, and only points close to $v = v_{\mathrm{peak}}$ contribute significantly. As a consequence, the first contribution (i), related to the effective length of the curve, becomes subdominant. The remaining dominant effect (ii) is then the intersection of the curve with the localized high-probability region at $v = v_{\mathrm{peak}}$. In this case, the dominant contribution corresponds to the curve tangent to the horizontal line $v = v_{\mathrm{peak}}$, which is associated with the second peak $\dot{\phi}_{\mathrm{peak,2}}$, with $\dot{\phi}_{\mathrm{peak,2}} > \dot{\phi}_{\mathrm{peak,1}}$. Therefore, the second peak $\dot{\phi}_{\mathrm{peak,2}}$ arises from the following contribution: (ii*) The tangent curve to the horizontal line $v = v_{\mathrm{peak}}$ in the $(\theta,v)$ plane, corresponding to the configuration that maximizes intersections with the localized high-probability region (yellow horizontal line).
   \item In the limiting case where the peculiar velocity distribution approaches a Dirac delta function, $P_v(v) \sim \delta(v - v_{\mathrm{peak}})$ (i.e.\ $\sigma_v \lesssim 10^{-3}$, see Fig~\ref{fig3}(a)), the most probable apparent angular velocity, $\dot{\phi}_{\mathrm{peak,2}}$, corresponds to the point where the curve $\gamma_{\dot{\phi}_{\mathrm{app}}}$ exhibits a local minimum at $(\theta_0, v_0 = v_{\mathrm{peak}})$ while being tangent to $v = v_{\mathrm{peak}}$. This is obtained by minimizing $\left. \frac{\partial g}{\partial \theta} \right|_{\theta = \theta_0} = 0$, which yields, in observational units of $\mathrm{mas}\,\mathrm{yr}^{-1}$,
   \begin{equation}
   \dot{\phi}_{\mathrm{peak,2}} = \frac{\frac{81}{125\pi}}{\chi(t_{\mathrm{emit}})} \frac{v_{\mathrm{peak}}}{\sqrt{1 - v_{\mathrm{peak}}^2}}.
   \label{eq31}
   \end{equation}
   In this regime, the apparent angular velocity PDF becomes sharply peaked (See the left panel in Fig~\ref{fig3}(a)) and then immediately falls to zero, since the curve $\gamma_{\dot{\phi}_{\mathrm{app}}}$ no longer intersects the horizontal line $v = v_{\mathrm{peak}}$ for large values of $\dot{\phi}_{\mathrm{app}} > \dot{\phi}_{\mathrm{peak,2}}$ (see the color plot in Fig~\ref{fig3}(a)).
 \begin{figure}[!htbp]
\centering
\includegraphics[width=0.9\textwidth]{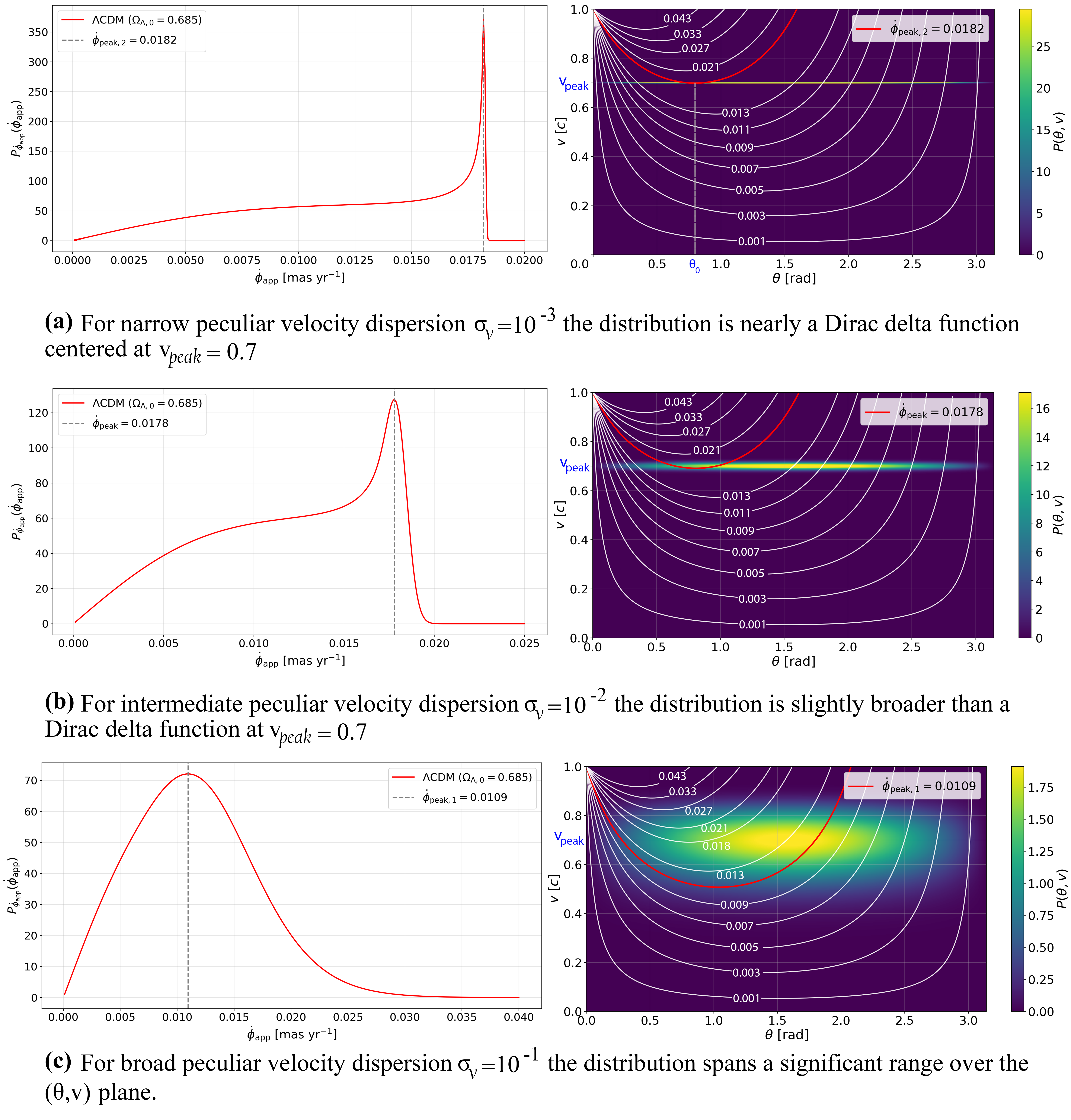}
\caption{\\[1ex]\textbf{Left panels:} Apparent angular velocity PDF, $P_{\dot{\phi}_{\rm app}}$, as a function of $\dot{\phi}_{\rm app}$ for a fixed population of luminous sources at redshift $z=1$ . The most probable peculiar velocity is set to $v_{\mathrm{peak}} = 0.7$. Results are shown for the $\Lambda$CDM cosmological model. \\[1ex]
\textbf{Right panels:} Color plots of the joint weight $P(\theta,v) = P_\theta(\theta)\, P_v(v)$ in the $(\theta,v)$ plane, with curves $\gamma_{\dot{\phi}_{\rm app}}$ indicating constant $\dot{\phi}_{\rm app}$. The red curve highlights the peak where $\dot{\phi}_{\rm app}$ reaches its maximum PDF value..}
\label{fig3}
\end{figure}
   \item Still in this limit, $\sigma_v \lesssim 10^{-3}$, the most probable value $\dot{\phi}_{\mathrm{peak,2}}$ given in Eq.~\eqref{eq31} provides a direct and robust observable, which is preferable to using the mean value $\langle \dot{\phi}_{\mathrm{app}} \rangle$ for distinguishing between different cosmological models. In this context, we can explicitly determine the most probable apparent angular velocity $\dot{\phi}_{\mathrm{peak,2}}$ for the two limiting cases of the cosmological constant: $\Omega_{\Lambda,0} = 0$ (MD) and $\Omega_{\Lambda,0} = 1$ ($\Lambda$D) by using the expression of $\chi(t_{\text{emit}})$ in Eq.~\eqref{eq29}, yielding
   \begin{subequations}
   \begin{equation}
   \dot{\phi}_{\rm peak,2}^{(\Omega_{\Lambda,0}=0)} = \frac{\frac{81}{125\pi} \, \frac{H_0}{2}}{1 - 1/\sqrt{1+z}} \, \frac{v_{\rm peak}}{\sqrt{1 - v_{\rm peak}^2}},
   \label{eq32a}
   \end{equation}
   \begin{equation}
   \dot{\phi}_{\rm peak,2}^{(\Omega_{\Lambda,0}=1)} = \frac{\frac{81}{125\pi}\, H_0}{z} \, \frac{v_{\rm peak}}{\sqrt{1 - v_{\rm peak}^2}}.
   \label{eq32b}
   \end{equation}
   \end{subequations}
   Although both expressions share the same kinematic dependence on the intrinsic velocity $v_{\rm peak}$, their redshift dependence differs substantially. Both decrease with increasing redshift $z$, as expected. To quantify the sensitivity to the cosmological density parameter, we define a redshift-dependent \textit{relative contrast} between the boundary cosmologies, providing a convenient indicator for discriminating between cosmological scenarios:
   \begin{equation}
   \frac{\dot{\phi}_{\rm peak,2}^{(\Omega_{\Lambda,0}=0)} - \dot{\phi}_{\rm peak,2}^{(\Omega_{\Lambda,0}=1)}}{\frac{1}{2} \left( \dot{\phi}_{\rm peak,2}^{(\Omega_{\Lambda,0}=0)} + \dot{\phi}_{\rm peak,2}^{(\Omega_{\Lambda,0}=1)} \right)} = 2 \frac{(z-2)\sqrt{1+z}+2}{(z+2)\sqrt{1+z}-2}.
   \label{eq33}
   \end{equation}
   This quantity is fully model-independent: it does not depend on the cosmological parameters or the most probable peculiar velocity $v_{\rm peak}$, but only on the redshift $z$. While the most probable angular velocities~\eqref{eq32a} and~\eqref{eq32b} decrease with increasing $z$, and their absolute difference also decreases, the relative contrast between the matter-dominated ($\Omega_{\Lambda,0} = 0$) and dark energy-dominated ($\Omega_{\Lambda,0} = 1$) models increases with $z$. This implies that measurements at higher redshift provide more efficient discrimination between different cosmological scenarios. In Fig.~\ref{fig4}, the relative contrast is plotted together with the most probable apparent velocities for both matter- and dark energy-dominated universes.
   \end{itemize}
    \begin{figure}[!htbp]
   \center
\includegraphics[width=0.7\textwidth]{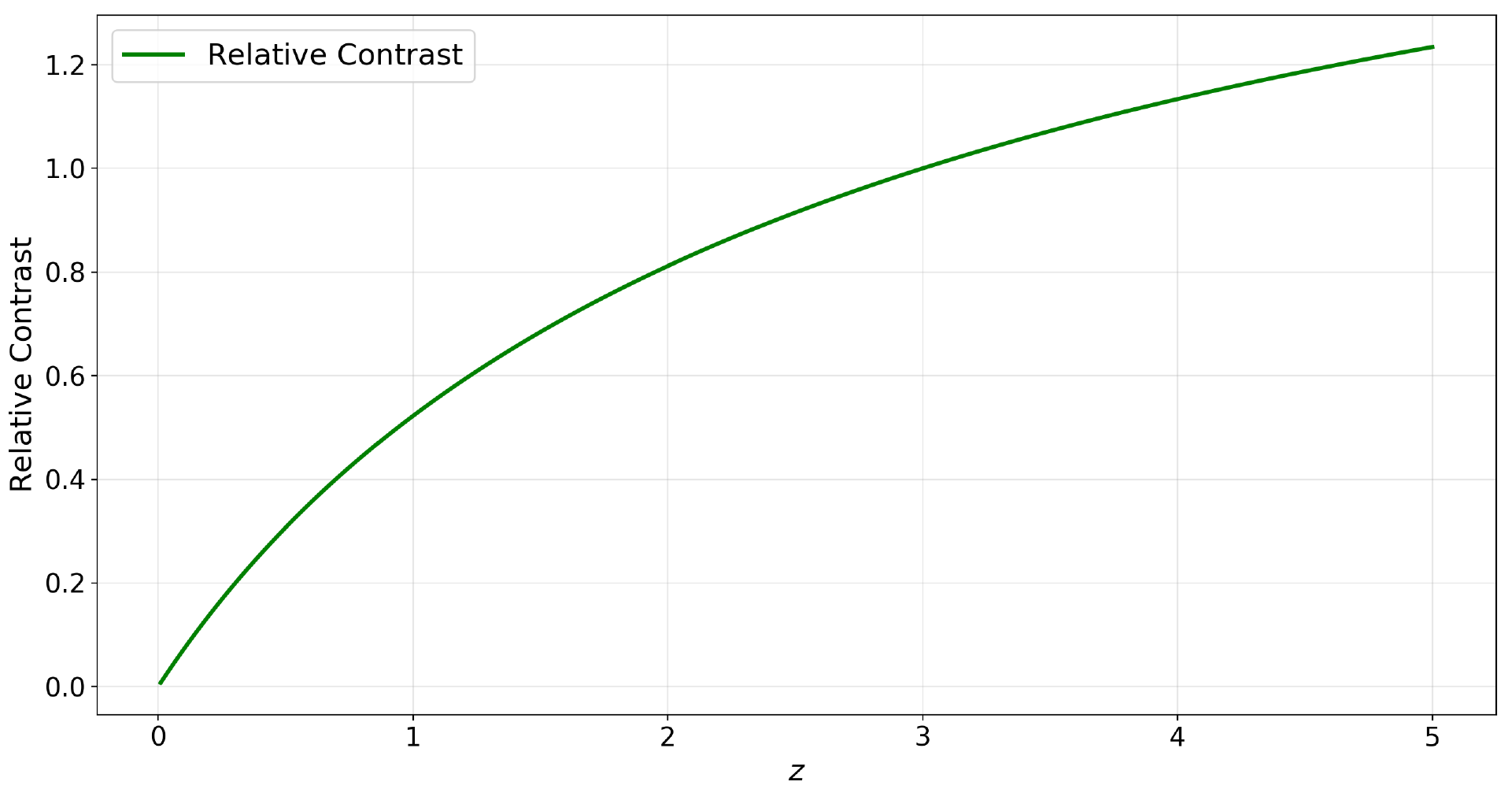}
    \caption{Illustrates the relative contrast as a function of redshift $z$, highlighting the ability to discriminate between different cosmological scenarios according to the dimensionless density parameter $\Omega_{\Lambda,0}$.}
    \label{fig4}
    \end{figure} 
\subsection{Apparent Transverse Velocity Distribution and Superluminal Fraction}

As discussed above, the physical apparent transverse velocity $v_{\text{app}}$ can be recovered from the observed angular velocity by multiplying by the angular diameter distance $D_A(z)$. This yields
\begin{equation}
v_{\text{app}} =
\frac{\frac{125\pi}{81}\chi\left(t_{\mathrm{emit}}\right)}
{1+z}
\,\dot{\phi}_{\mathrm{app}}^{\text{mas}\,\text{yr}^{-1}}
\label{eq34}.
\end{equation}
Therefore, the condition for apparent superluminal motion, $v_{\text{app}} > 1$, is equivalent to
\begin{equation}
\dot{\phi}_{\mathrm{app}}^{\text{mas}\,\text{yr}^{-1}}
>
\frac{1+z}
{\frac{125\pi}{81}\chi\left(t_{\mathrm{emit}}\right)}
\label{eq35}.
\end{equation}
The fraction of apparent superluminal sources, denoted $P(v_{\text{app}} > 1)$, can thus be obtained by integrating the probability density function given in Eq.~\eqref{eq30} over the domain $(\dot{\phi}_{\text{min}}, +\infty)$,
\begin{equation}
P\left(v_{\mathrm{app}}>1\right)
=
\int_{\dot{\phi}_{\mathrm{min}}}^{+\infty}
P_{\dot{\phi}_{\mathrm{app}}}
\left(
\dot{\phi}_{\mathrm{app}}
\right)
\, d\dot{\phi}_{\mathrm{app}}
\label{eq36},
\end{equation}
where
\begin{equation}
\dot{\phi}_{\text{min}}
=
\frac{1+z}
{\frac{125\pi}{81}\chi\left(t_{\mathrm{emit}}\right)}
\label{eq37},
\end{equation}
represents the minimum apparent angular velocity required for superluminal motion. This superluminal fraction depends on the cosmological parameters $H_0$ and $\Omega_{\Lambda,0}$, the redshift $z$, and the parameters of the intrinsic velocity distribution, namely $v_{\text{peak}}$ and $\sigma_v$. Alternatively, the superluminal fraction can also be determined by integrating the apparent transverse velocity PDF $P_{v_{\mathrm{app}}}\left(v_{\mathrm{app}}\right)$ over the domain $(1,+\infty)$. The apparent transverse velocity PDF is related to the apparent angular velocity PDF through the Jacobian transformation,
\begin{equation}
P_{v_{\mathrm{app}}}\left(v_{\mathrm{app}}\right)
=
\frac{1+z}
{\frac{125\pi}{81}\chi\left(t_{\mathrm{emit}}\right)}
P_{\dot{\phi}_{\mathrm{app}}}
\left(
\frac{1+z}
{\frac{125\pi}{81}\chi\left(t_{\mathrm{emit}}\right)}
v_{\mathrm{app}}
\right)
\label{eq38}.
\end{equation}
Therefore, the superluminal fraction can equivalently be written as
\begin{equation}
P\left(v_{\mathrm{app}}>1\right)
=
\int_{1}^{+\infty}
P_{v_{\mathrm{app}}}\left(v_{\mathrm{app}}\right)
\, dv_{\mathrm{app}}
\label{eq39}.
\end{equation}
It is worth noting from Eq.~\eqref{eq2} that the apparent transverse velocity $v_{\mathrm{app}}$ is affected by cosmic expansion only through the factor $a(t_{\mathrm{emit}})/a(t_{\mathrm{obs}}) = (1+z)^{-1}$, and therefore depends only on the redshift. Consequently, for a population of luminous sources located at a fixed redshift, the apparent transverse velocity PDF $P_{v_{\mathrm{app}}}\left(v_{\mathrm{app}}\right)$ is expected to be independent of the cosmological parameters $\Omega_{\Lambda,0}$ and $H_0$. In this case, all cosmological models predict the same apparent transverse velocity PDF for a source population observed at the same redshift.

In the limit of small dispersion $\sigma_v$, where the velocity distribution approaches a Dirac delta function $P_v(v) \simeq \delta(v - v_{\text{peak}})$, the superluminal fraction is non-zero only if $\dot{\phi}_{\text{min}} < \dot{\phi}_{\text{peak,2}}$, where $\dot{\phi}_{\text{peak,2}}$, given in Eq.~\eqref{eq31}, corresponds to the maximum attainable apparent angular velocity. This condition leads to the following constraint on the intrinsic velocity,
\begin{equation}
v_{\text{peak}}
>
\frac{1+z}{\sqrt{1+(1+z)^2}}
\label{eq40}.
\end{equation}
This relation defines the minimum intrinsic velocity required for a population at redshift $z$ to exhibit apparent superluminal motion in the limit of a sharply peaked velocity distribution. In the low-redshift limit $z \to 0$, this reduces to $v_{\text{peak}} > \frac{1}{\sqrt{2}}$, which is the well-known condition for the occurrence of superluminal motion in the local (low-redshift) regime. Under this limit, and together with the condition in Eq.~\eqref{eq40} defining the intrinsic velocity threshold for apparent superluminal motion at redshift $z$, the superluminal fraction can be computed by integrating the apparent angular velocity PDF in Eq.~\eqref{eq30} over the bounded interval $(\dot{\phi}_{\text{min}}, \dot{\phi}_{\text{peak,2}})$.

Finally, a numerical evaluation of the superluminal fraction $P(v_{\text{app}} > 1)$, assuming a Dirac delta velocity distribution $P_v(v) = \delta(v - v_{\text{peak}})$ and an isotropic angular distribution $P_\theta(\theta) = \frac{1}{2}\sin\theta$ for a population of luminous sources at $z=1$, gives the minimum intrinsic velocity required for superluminal motion, $\frac{2}{\sqrt{5}} \approx 0.89$. Therefore, apparent superluminal motion can occur only for $v_{\text{peak}} \gtrsim 0.89$. The corresponding superluminal fractions for different values of $v_{\text{peak}}$ are approximately: $0\%$ for $v_{\text{peak}} = 0.7$, $0\%$ for $v_{\text{peak}} = 0.8$, $\sim 5\%$ for $v_{\text{peak}} = 0.9$, $\sim 19\%$ for $v_{\text{peak}} = 0.99$, and $\sim 20\%$ for $v_{\text{peak}} = 0.999$. In the limit $v_{\text{peak}} \rightarrow 1$, the superluminal fraction tends toward approximately $20\%$. Importantly, these fractions do not represent realistic observational estimates, since the present calculation assumes a purely isotropic angular distribution. In practice, observations are strongly biased toward luminous sources with small viewing angles relative to the line of sight. In the context of relativistic jets, such systems correspond to blazars, where Doppler boosting plays a major observational role. A more realistic treatment including Doppler boosting effects, physically motivated angular and velocity distributions, and comparison with available observational data will be investigated in future work.
  \section{Constraining Cosmological Parameters From Apparent Angular Velocity PDF}\label{section5}
From Fig.~\ref{fig2}, it is evident that the apparent angular velocity PDF corresponding to different cosmological parameter pairs $(\Omega_{\Lambda,0},H_0)$ can substantially overlap. In particular, distinct parameter sets $(\Omega_{\Lambda,0}^{(1)},H_0^{(1)})$ and $(\Omega_{\Lambda,0}^{(2)},H_0^{(2)})$ may produce identical apparent angular velocity distributions. This degeneracy arises from the non--injective mapping $(\Omega_{\Lambda,0},H_0)\mapsto \chi(t_{\mathrm{emit}})$ in Eq.~\eqref{eq29}: different combinations of $\Omega_{\Lambda,0}$ and $H_0$ can yield the same value of the comoving distance at the emission time, and therefore lead to indistinguishable apparent angular velocity PDFs.
\begin{figure}[!htbp]
\centering
\includegraphics[width=0.7\textwidth]{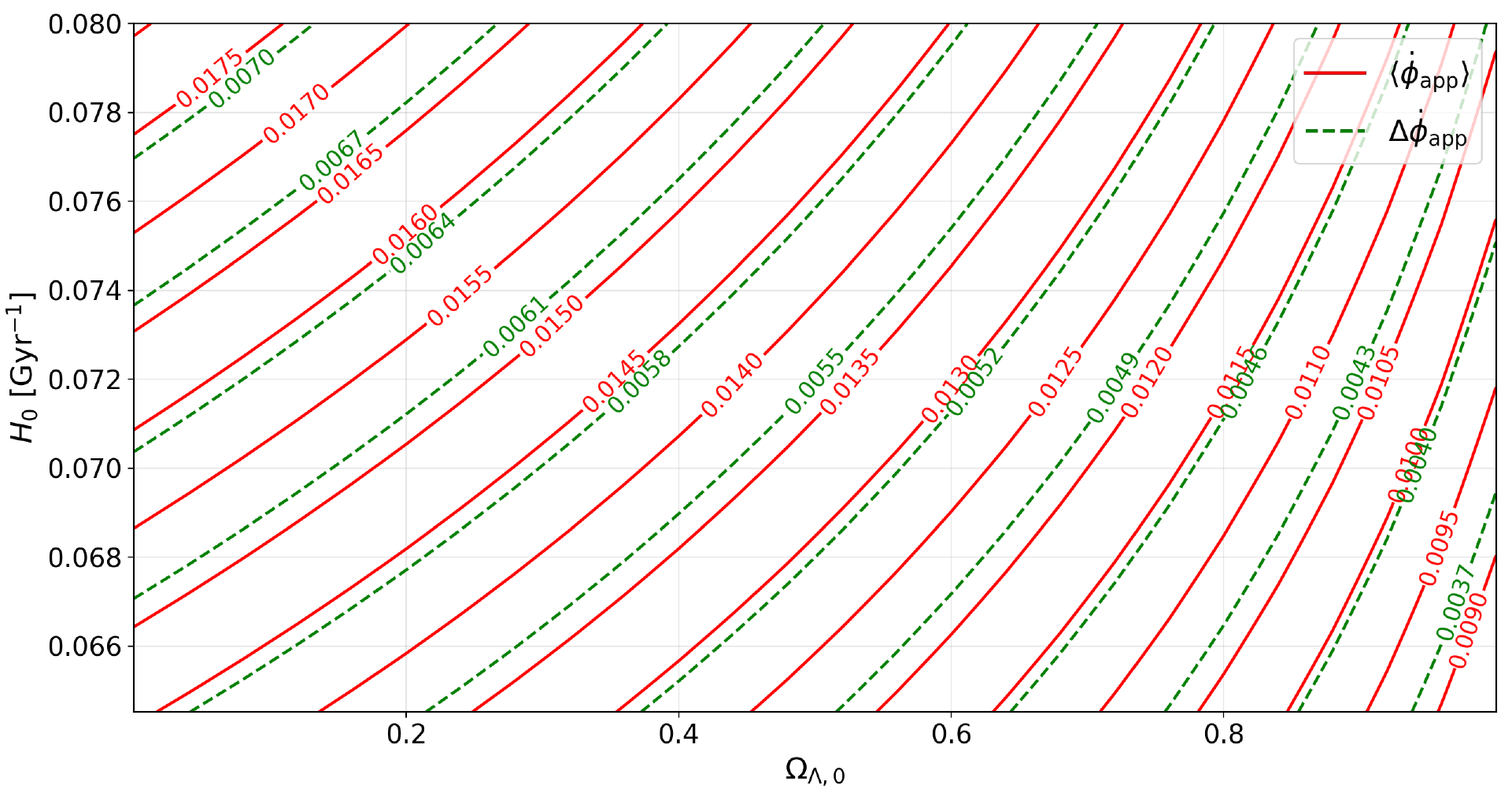}
\caption{Implicit curves of $\langle \dot{\phi}_{\mathrm{app}} \rangle(\Omega_{\Lambda,0},H_0)$ (red) and $\Delta \dot{\phi}_{\mathrm{app}}(\Omega_{\Lambda,0},H_0)$ (dashed green) for a set of constant values. Both sets of curves assume a most probable peculiar velocity $v_{\rm peak}=0.7$, a logit-normal dispersion $\sigma_v = 10^{-2}$, and a source redshift $z=1$.}
\label{fig5}
\end{figure} 
Consequently, if the goal is to constrain both $\Omega_{\Lambda,0}$ and $H_0$ simultaneously, a statistical modeling of apparent transverse velocities for a population of luminous sources located at a fixed redshift is, by itself, insufficient. At best, the present framework provides a correlation between $\Omega_{\Lambda,0}$ and $H_0$, rather than unique constraints on each parameter independently. Since all observables ($O$) derived from this effect depend on $(\Omega_{\Lambda,0},H_0)$ only through the comoving distance $\chi(t_{\mathrm{emit}})$ (see Eq.~\eqref{eq30}), contours of different observables, $O(\Omega_{\Lambda,0},H_0)=O_{\mathrm{obs}}$ form parallel curves in the $(\Omega_{\Lambda,0},H_0)$ parameter space. For example, observables such as the mean apparent angular velocity $\langle \dot{\phi}_{\mathrm{app}} \rangle$ and its standard deviation $\Delta \dot{\phi}_{\mathrm{app}}$ define degenerate constraint curves $\langle \dot{\phi}_{\mathrm{app}} \rangle(\Omega_{\Lambda,0},H_0)=\langle \dot{\phi}_{\mathrm{app}} \rangle_{\mathrm{obs}}$ and $\Delta \dot{\phi}_{\mathrm{app}}(\Omega_{\Lambda,0},H_0)=\Delta v_{\mathrm{app,obs}}$, which provide correlated rather than independent constraints on $\Omega_{\Lambda,0}$ and $H_0$, with uncertainties determined by observational errors. This behavior is illustrated in Fig.~\ref{fig5}, where contours of constant $\langle \dot{\phi}_{\mathrm{app}} \rangle$ and constant $\Delta \dot{\phi}_{\mathrm{app}}$ in the $(\Omega_{\Lambda,0},H_0)$ plane are shown to be parallel, confirming the degeneracy discussed above.

\begin{figure}[!htbp]
\centering
\includegraphics[width=0.7\textwidth]{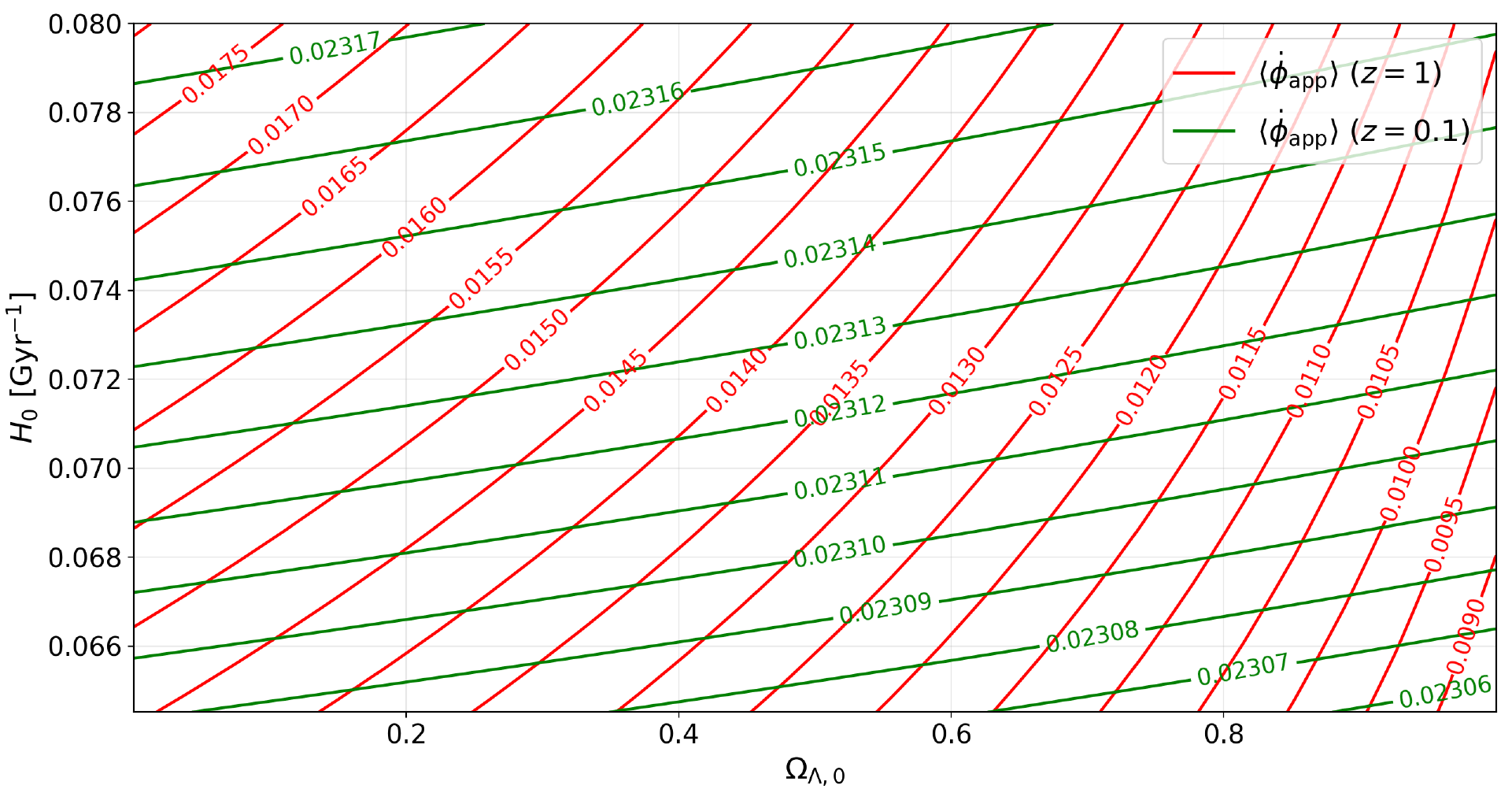}
\caption{Implicit curves of $\langle \dot{\phi}^{(z=1)}_{\mathrm{app}} \rangle(\Omega_{\Lambda,0},H_0)$ ($z=1$; red) and $\langle \dot{\phi}^{(z=0.1)}_{\mathrm{app}} \rangle(\Omega_{\Lambda,0},H_0)$ ($z=0.1$; green)
        for a set of constant values. Both sets of curves assume a most probable peculiar velocity $v_{\rm peak}=0.7$, and a logit-normal dispersion $\sigma_v = 10^{-2}$.}
\label{fig6}
\end{figure}

To overcome this degeneracy, one can exploit the fact that the orientation of these degeneracy directions depends on the redshift of the source population. Consequently, rather than relying on multiple independent observables to determine the cosmological parameters $(\Omega_{\Lambda,0},H_0)$, we propose using a single observable measured for two distinct source populations located at different redshifts. This approach yields non--parallel constraint contours in the $(\Omega_{\Lambda,0},H_0)$ parameter space. Concretely, consider the mean apparent angular velocity $\langle \dot{\phi}_{\mathrm{app}} \rangle$ measured for two populations at redshifts $z=1$ and $z=0.1$. The corresponding constraints,
\begin{subequations}
\begin{align}
\langle \dot{\phi}_{\mathrm{app}}^{(z=1)} \rangle(\Omega_{\Lambda,0},H_0)=\langle \dot{\phi}_{\mathrm{app}}^{(z=1)} \rangle_{\mathrm{obs}}
\label{eq41a},
\end{align}
\begin{align}
\langle \dot{\phi}_{\mathrm{app}}^{(z=0.1)} \rangle(\Omega_{\Lambda,0},H_0)=\langle \dot{\phi}_{\mathrm{app}}^{(z=0.1)} \rangle_{\mathrm{obs}},
\label{eq41b}
\end{align}
\end{subequations}
 define two implicit sets of curves in the $(\Omega_{\Lambda,0},H_0)$ plane that are generically non--parallel. Their intersection therefore allows both $\Omega_{\Lambda,0}$ and $H_0$ to be constrained simultaneously, within the observational error bars. This idea is examined numerically by plotting the implicit curves defined by Eqs.~\eqref{eq41a} and~\eqref{eq41b} in the $(\Omega_{\Lambda,0},H_0)$ plane, using the same parameter values adopted in the previous figures. The results are shown in Fig.~\ref{fig6}, which demonstrates that combining measurements from populations at different redshifts offers a viable approach to probing the underlying cosmological model, without requiring direct distance measurements to the sources. 
  Finally, this method can be extended to more general cosmological models that include not only matter and dark energy, but also radiation, curvature (by using the appropriate metric distance), or even models with time-varying dark energy density (e.g., $\text{w}_0\text{w}_a$CDM model). In fact, any cosmological model characterized by $n$ degrees of freedom parameters requires an $n$-dimensional statistical analysis across $n$ different cosmological redshifts.
\subsection*{Error Analysis and Observational Requirements for Cosmological Constraints}
We now ask how many independent sources $N$, located at approximately the same redshift, are required to constrain cosmological parameters with an accuracy comparable to current observations. We also estimate the angular resolution required for detectors to measure the apparent angular velocities with sufficiently small systematic errors, denoted $(\Delta \dot{\phi}_{\mathrm{app}})_{\mathrm{SYS}}$. It is worth noting that the latest value of the Hubble constant is not unique due to the well-known Hubble tension~\cite{Wong2020,Verde2019}. Two main high-precision determinations exist: Early-Universe (EU)~\cite{Aghanim2020}, $H_0 = 67.4 \pm 0.5 \ \mathrm{km \ s^{-1} \ Mpc^{-1}}$, and Late-Universe (LU)~\cite{Riess2016,Riess2022}, $H_0 = 73.0 \pm 1.0 \ \mathrm{km \ s^{-1} \ Mpc^{-1}}$. Expressed in $\mathrm{Gyr}^{-1}$, the corresponding uncertainties are approximately $(\Delta H_0)_{\mathrm{EU}} \sim 10^{-4} \ \mathrm{Gyr}^{-1}$ and $(\Delta H_0)_{\mathrm{LU}} \sim 10^{-3} \ \mathrm{Gyr}^{-1}$. In the following discussion, we adopt the more stringent value $\Delta H_0 \sim 10^{-4} \ \mathrm{Gyr}^{-1}$. For the dimensionless dark energy density, we use the Planck 2018 result~\cite{Aghanim2020} $\Omega_{\Lambda,0} = 0.684 \pm 0.007$, and therefore take $\Delta \Omega_{\Lambda,0} \sim 10^{-3}$.
For simplicity, we adopt a linear approximation and evaluate derivatives using peak values within the $\Lambda$CDM model with parameters $(\Omega_{\Lambda,0} = 0.685, H_0 = 1/14.5 \ \mathrm{Gyr}^{-1})$. This approximation allows us to estimate the order of magnitude of systematic errors and the required number of independent sources. Assuming a small peculiar velocity dispersion $\sigma_v = 10^{-3}$ and a redshift $z=1$.
\\

Observational uncertainties can be divided into systematic errors due to instrumental resolution, $(\Delta \dot{\phi}_{\mathrm{app}})_{\mathrm{SYS}}$, and statistical errors of the mean, $(\Delta \langle \dot{\phi}_{\mathrm{app}} \rangle)_{\mathrm{Stat}} = \Delta \dot{\phi}_{\mathrm{app}}/\sqrt{N}$. This expression remains valid even if the distribution $P_{\dot{\phi}_{\mathrm{app}}}(\dot{\phi}_{\mathrm{app}})$ is non-Gaussian as ensured by the Central Limit Theorem (CLT), where $\Delta \dot{\phi}_{\mathrm{app}}$ represents the standard deviation of the $\Lambda$CDM distribution, estimated from Fig.~\ref{fig2} as $\Delta \dot{\phi}_{\mathrm{app}} \sim 10^{-3} \ \mathrm{mas \ yr^{-1}}$. Both systematic and statistical uncertainties must be smaller than the propagated errors induced by cosmological parameters, namely $(\Delta \langle \dot{\phi}_{\mathrm{app}} \rangle)_{\Omega_{\Lambda,0}}$ and $(\Delta \langle \dot{\phi}_{\mathrm{app}} \rangle)_{H_0}$, which arise from the independent uncertainties $\Delta \Omega_{\Lambda,0}$ and $\Delta H_0$, respectively. To first order, these contributions can be estimated as:
\begin{subequations}
\begin{align}
(\Delta \langle \dot{\phi}_{\mathrm{app}} \rangle)_{\Omega_{\Lambda,0}} \approx \left| \frac{\partial \langle \dot{\phi}_{\mathrm{app}} \rangle}{\partial \Omega_{\Lambda,0}} \right| \Delta \Omega_{\Lambda,0}
\label{eq42a},
\end{align}
\begin{align}
(\Delta \langle \dot{\phi}_{\mathrm{app}} \rangle)_{H_0} \approx \left| \frac{\partial \langle \dot{\phi}_{\mathrm{app}} \rangle}{\partial H_0} \right| \Delta H_0
\label{eq42b}.
\end{align}
\end{subequations}
Using linear approximation in $\Omega_{\Lambda,0}$ and $H_0$, these derivatives can be estimated either by visual inspection of Fig.~\ref{fig2} or through numerical evaluation of Eqs.~\eqref{eq32a} and~\eqref{eq32b}. The full variation over the range $\Delta \Omega_{\Lambda,0} = 1$, corresponding to:
\begin{equation}
\left| \frac{\partial \langle \dot{\phi}_{\mathrm{app}} \rangle}{\partial \Omega_{\Lambda,0}} \right| \approx \dot{\phi}_{\mathrm{peak,2}}^{(\Omega_{\Lambda,0}=0, H_0=1/14.5)} - \dot{\phi}_{\mathrm{peak,2}}^{(\Omega_{\Lambda,0}=1, H_0=1/14.5)} \sim 10^{-2} \ \mathrm{mas \ yr^{-1}}
\label{eq43},
\end{equation}
based on the separation between the peaks of the green ($\Omega_{\Lambda,0}=0$) and blue ($\Omega_{\Lambda,0}=1$) curves in Fig.~\ref{fig2}. This leads to an estimated contribution from the dark energy density uncertainty of the order:
\begin{equation}
(\Delta \langle \dot{\phi}_{\mathrm{app}} \rangle)_{\Omega_{\Lambda,0}} \sim 10^{-2} \ \mu\mathrm{as \ yr^{-1}}
\label{eq44}.
\end{equation}
Similarly, for $H_0$, we approximate:
\begin{equation}
\left| \frac{\partial \langle \dot{\phi}_{\mathrm{app}} \rangle}{\partial H_0} \right| \approx \frac{\dot{\phi}_{\mathrm{peak,2}}^{(\Omega_{\Lambda,0}=0.685, H_0=1/12.5)} - \dot{\phi}_{\mathrm{peak,2}}^{(\Omega_{\Lambda,0}=0.685, H_0=1/14.5)}}{(1/12.5 - 1/14.5)} \sim 10^{-1} \ \mathrm{mas \ yr^{-1} \ Gyr}
\label{eq45},
\end{equation}
from the difference between the peaks of the dashed red ($H_0=1/12.5$) and red ($H_0=1/14.5$) curves in Fig.~\ref{fig2}, yielding:
\begin{equation}
(\Delta \langle \dot{\phi}_{\mathrm{app}} \rangle)_{H_0} \sim 10^{-2} \ \mu\mathrm{as \ yr^{-1}}
\label{eq46}.
\end{equation}
Therefore, to achieve constraints at the level $\Delta \Omega_{\Lambda,0} \sim 10^{-3}$ and $\Delta H_0 \sim 10^{-4} \ \mathrm{Gyr}^{-1}$, the systematic and statistical uncertainties must satisfy:
\begin{subequations}
\begin{align}
(\Delta \dot{\phi}_{\mathrm{app}})_{\mathrm{SYS}} \lesssim 10^{-2} \ \mu\mathrm{as \ yr^{-1}}
\label{eq47a},
\end{align}
\begin{align}
(\Delta \langle \dot{\phi}_{\mathrm{app}} \rangle)_{\mathrm{Stat}} \lesssim 10^{-2} \ \mu\mathrm{as \ yr^{-1}}
\label{eq47b}.
\end{align}
\end{subequations}
This requirement implies a minimum number of independent sources $N \gtrsim 10^4$ at approximately the same redshift. Additionally, the angular resolution required for detectors is of order $\sim 10^{-2}\,\mu\mathrm{as\,yr^{-1}}$. Alternatively, long-term monitoring over a time baseline of $\sim 10$ years can relax this requirement, leading to an effective sensitivity of $\sim 10^{-1}\,\mu\mathrm{as}$ over a decade.
\\

Encouragingly, current VLBI systems already achieve astrometric precision at the level of $\sim 10\,\mu\mathrm{as}$, with the best cases approaching $\sim 1\,\mu\mathrm{as}$. The next generation of instruments, particularly Square Kilometre Array--enabled VLBI (SKA-VLBI), is expected to routinely push this precision below $10\,\mu\mathrm{as}$ and toward the $\sim 1\,\mu\mathrm{as}$ regime through major gains in sensitivity, wider bandwidths, and advanced calibration techniques~\cite{Li2024,Reid2014}. Additional progress will come from arrays such as the next-generation Event Horizon Telescope and the ngVLA, which will improve both temporal sampling and image fidelity, allowing more precise tracking of compact structures. In the longer term, space-based VLBI missions such as \textit{Millimetron} will extend baselines far beyond Earth’s diameter, potentially reaching sub-microarcsecond ($\sim 10^{-1}\,\mu\mathrm{as}$) precision. Despite these impressive advances, the required sensitivity of $\sim 10^{-2}\,\mu\mathrm{as\,yr^{-1}}$ remains about one to two orders of magnitude below the best currently achieved astrometric precision. Even when considering decade-long baselines, which effectively bring the requirement to $\sim 10^{-1}\,\mu\mathrm{as}$, this target lies at the very edge of the capabilities anticipated for future space-VLBI missions~\cite{Johnson2023,Gurvits2020}. Therefore, while upcoming facilities significantly narrow the gap, reaching the precision needed to detect such extremely small angular drifts remains a challenging objective for the next generation of instruments.
\\

On the other hand, current VLBI monitoring programs have already made substantial progress in building large statistical samples. Observations of relativistic jets in active galactic nuclei now include several hundred sources, as demonstrated by long-term surveys such as the MOJAVE program ($\sim 437 AGNs$) ~\cite{Lister2019b} and earlier VLBI surveys like the CJF sample ($\sim 300$ sources) ~\cite{Britzen2008}. Complementary programs such as TANAMI extend these studies to the southern hemisphere, monitoring a smaller but well-defined sample of $\sim 80$ AGN jets~\cite{Muller2013}. VLBI monitoring programs track multiple components per jet across several hundred AGNs, typically identifying a few components per source and measuring apparent angular motions on the order of milliarcseconds per year~\cite{Reid2014,Lister2019b}. Although this represents a significant dataset, it still falls short of the requirement $N \gtrsim 10^4$ by roughly an order of magnitude. However, forthcoming facilities such as SKA-VLBI, the ngVLA, and the next-generation Event Horizon Telescope are expected to dramatically increase both the number of detected jets and the number of trackable components, potentially reaching samples of several thousands to tens of thousands of sources. In this sense, while the present-day observations do not yet satisfy the required statistics, the expected growth in survey size and sensitivity suggests that this condition may become achievable in the near future, especially when combined with improved detection of fainter and more distant populations~\cite{Paragi2015,Murphy2018,Johnson2023,Kardashev2014,Gurvits2020}.
\section{Conclusion}\label{section6}
In this work, we have employed a self-consistent statistical method to study apparent superluminal motions of relativistic sources in an expanding universe, with explicit inclusion of cosmological effects through the comoving distance at the emission time. Starting from exact relativistic kinematics and a minimal set of population assumptions, we derived the probability distributions of the apparent angular velocity (PDF) for a population of luminous sources at a fixed redshift $z$ and investigated their sensitivity to the underlying cosmological parameters.

we showed that both the mean value and the width of the apparent angular velocity distribution respond systematically to variations in the dark energy density parameter $\Omega_{\Lambda,0}$ and the Hubble parameter $H_0$. These dependencies allow apparent angular velocity statistics to be used as cosmological observables within a controlled statistical framework. 

In the limiting regime of an extremely narrow peculiar velocity distribution, where $P_v(v)$ approaches a Dirac delta function, we identified the most probable apparent angular velocity, $\dot{\phi}_{\mathrm{peak,2}}$, as a particularly robust and well-defined observable. In this case, $\dot{\phi}_{\mathrm{peak,2}}$ admits a closed-form expression that is inversely proportional to the comoving distance at emission and depends only on the most probable intrinsic velocity $v_{\mathrm{peak}}$ and the Hubble parameter $H_0$. Remarkably, the relative contrast between matter- and dark energy–dominated cosmologies is independent of both $H_0$ and $v_{\mathrm{peak}}$, depending only on the redshift $z$. This property makes $\dot{\phi}_{\mathrm{peak,2}}$ a powerful discriminator between cosmological scenarios, particularly at high redshift where the contrast is maximized.

Moreover, we constructed implicit constraints in the $(\Omega_{\Lambda,0},H_0)$ parameter space using statistical observables and found that apparent angular velocity statistics generally define correlated degeneracy directions rather than unique parameter values, since different combinations of $\Omega_{\Lambda,0}$ and $H_0$ can produce identical values of the cosmoving distance at the emission time $\chi(t_\text{emit})$, leading to parallel constant--observable curves in the $(\Omega_{\Lambda,0},H_0)$ plane. Importantly, we found that the orientation of these degeneracy directions depends on the redshift $z$ of the source population. As a consequence, combining measurements of the same observable (e.g., the mean value) for two distinct source populations at different redshifts can lift the degeneracy and enable a systematic exploration of the $(\Omega_{\Lambda,0},H_0)$ parameter space. In this sense, apparent superluminal motion provides a complementary kinematic consistency test for cosmological models without a distance ladder, rather than a standalone precision probe.

In addition, the error analysis presented in this work shows that achieving competitive cosmological constraints with apparent angular velocity measurements requires both ultra-high astrometric precision and large statistical samples. In particular, sensitivities at the level of $\sim 10^{-2}\,\mu\mathrm{as\,yr^{-1}}$ and source counts $N \gtrsim 10^4$ are needed to match current constraints on $H_0$ and $\Omega_{\Lambda,0}$. While current VLBI capabilities remain short of these requirements, upcoming facilities are expected to significantly reduce the gap in both precision and sample size. This highlights the strong potential of future VLBI observations as an independent cosmological probe.

Overall, this study clarifies the potential of using relativistic jet kinematics for cosmological inference. Apparent angular velocity statistics can determine cosmological parameters, they encode non-trivial information about the expansion history when analyzed within a controlled statistical framework. The framework introduced here is readily extendable to broader cosmological scenarios (including curvature, radiation, or time-varying dark energy density), where an $n$-parameter model can, in principle, be tested through an $n$-dimensional statistical analysis across $n$ redshift slices. Future extensions incorporating realistic source populations, redshift evolution, and observational selection functions will be necessary to connect this framework to real data and to assess its full potential in conjunction with existing cosmological probes. In particular, improvements in measuring the smallest possible angular displacements using very long baseline interferometry (VLBI) will enhance the precision of apparent angular velocity measurements and strengthen the resulting cosmological constraints.

\end{document}